# Angular Momentum Operators from Quantized $SO(3)$

Ahmad Adel Abutaleb

Department of Mathematics, Faculty of Science, University of Mansoura, Elmansoura 35516, Egypt.

## Abstract

In this paper, we will assume that the structure picture of the rotation angles will be changed according to the scale of measurement (minimum measurable angle) and if we have a device with very high accuracy (high resolution) then we can notice a discrete nature of the rotations. We derived the form of the angular momentum matrices and angular momentum operators in this case and we find an indication of the need to change all quantum mechanical operators at this very small scale (high energy level). As a physical consequence, we calculated the magnetic quantum number and find that it has been shifted to a fractional multiples of $h$ and therefore the spin of quantum particles is no longer take integer or half integer values but some fractional values between them.

## Introduction

In 1988, Stefan Hilger in his Ph.D thesis established a new branch of mathematics called time-scales calculus in order to unify continuous and discrete analysis (differential and difference equations) [1]. The general idea of time-scale calculus is to study the dynamic equations where the domain of the functions is an arbitrary closed subset of the real numbers called time-scale $T$ [2-6]. When $T = R$ (the real numbers), we recover our usual frame work of calculus, and when $T = Z$ (the integer numbers), we arrive at the usual version of difference equations. This new branch of mathematics generalizes the results of ordinary calculus in a revolutionary way. For example, consider a function $f: D \to R$ and suppose that there is some point of discontinuity $x \in D$. In our ordinary calculus, we cannot speak about the differentiability of the function $f$ at , but in the time scale calculus we can differentiate the function $f$ at any point even the function $f$ was discontinuous (in ordinary sense) at every point of its domain. Moreover, we can study the dynamic equations even the domain $D$ has a fractal nature. (notice that in a fractal-like curve, the concept of the neighborhood was destroyed (in Euclidean Topology) Because the distance of any tow points on this curve is infinity) [7]. Some results of physical literatures expect a kind of discretization nature of flat [8-10] and curved Spacetime [11], so the questions about the form of the theories in discrete spacetime are natural. There is a huge amount of mathematical works on time scale which may be very useful for physicists if they wish to study and analyze the form of physical theories (Special relativity, General relativity, Electromagnetic,…etc) on a general structure of the spacetime. e.g. [12-16]. Roughly speaking, by using time-scale calculus, the usual physical theories seems to be special cases of more general theories. As a simple example of the importance of time-scale calculus, we derived the form of the angular momentum matrices and angular momentum operators in usual continuous smooth spacetime but with assumption that the rotation angles around any axis can only take discrete values. A substantial difference appears only when the very small quantities should be taken into account. If we assume the quantization of the space itself, then the rotation angles are already quantized, and therefore we have an indication of the need to change all quantum mechanical operators at very small scale (high energy level). Finally, we derived the values of quantum magnetic number in the low scale.

## Preliminaries

A time scale $T$ is an arbitrary closed subset of the real numbers. The forward $\sigma$ and backward $\rho$ operators defined by

$\sigma(t):= \text{Inf}\{s \in T, s > t\}$ , $\rho(t):= \text{Sup}\{s \in T, s < t\}$.

A point $t \in T$ is said to be right dense if $\sigma(t) = t$, left dense if $\rho(t) = t$, right scattered if $\sigma(t) > t$, left scattered if $\rho(t) < t$.

The graininess function $\mu$ is defined by $\mu(t):= \sigma(t) - t$ .

A function $f: T \to R$ is said to be rd-continuous if it is continuous at each right dense points and if the left limit exist for each left dense points.

The function $f$ is said to be regulated provided its right sided limits exist at all right dense points in $T$ and its left sided limits exist at all left dense points in $T$.

The Hilger derivative of $f$ at $t \in T$ is defined by
$f^\Delta(t) = \lim_{s \to t} \frac{f(\sigma(t)-f(s))}{\sigma(t)-s}$, $s \in U(t)$, where, $U(t) = T - \{\sigma(t)\}$ provided that limit exists.

A continuous function $f: T \to R$ is called pre-differentiable with region of differentiation $D$, provided $D \subset T^k$, $T^k - D$ is countable and contains no right scattered points of $T$, and $f$ is differentiable at each point $t \in D$. Notice that if $T$ has a left scattered maximum $m$ then $T^k = T - \{m\}$ other wise $T^k = T$.

**Theorem**

If $f$ is a regulated function, then there exist a function $F$ which is a pre differentiation with region of differentiation , such that $F^\Delta(t) = f(t)$ for all $t \in D$.
Assume $f: T \to R$ is a regulated function. Any function $F$ as in the last theorem is called a pre anti derivative of $f$. the indefinite integral of a regulated function $f$ is defined as
$\int f(t) \Delta t = F(t) + c$, $c$ is an arbitrary constant.

**What do we mean by measurement scale?**

We define the measurement scale of any physical quantity as the minimum possible observed value of this quantity. Therefore our observation regarding the structure of the space will be change according to the change of this scale. For example, suppose that there is some Electron moving along $x$ axis with the following possible discrete values,

..., $-2 \times 10^{-10}$, $-10^{-10}$, $0$, $10^{-10}$, $2 \times 10^{-10}$, ... . Suppose that we have two observers $A$ and $B$. $A$ is an observer with length scale= $10^{-9}$ (i.e. the minimum length which $A$ can measure is equal to $10^{-9}$) and $B$ is an observer with length scale= $10^{-11}$. If we ask the observer $A$ about the allowed positions of the Electron, then he will tell us that the Electron can take any position on $x$ axis (i.e. he see continuous moving of the electron) and the space will appear to him as a continuous space. On the other hand, observer $B$ will see discrete space, so the structure of space itself will change according to the scale of measurement.

Now, define $\theta_i$ as the rotation angle around $x^i$ and suppose that the rotations can only take discrete values. Consider some very large integer number $N$ (e.g. $= 10^{36}$) and take $l = \frac{2\pi}{N}$, and consider that the rotation angle around each axis can take only the following discrete values, $\theta_i \in \{0, l, 2l, \ldots \ldots, Nl\}$, $i = 1,2,3$.

The set of values $\theta \in T = \{0, l, 2l, \ldots \ldots, Nl\}$ is a closed subset of the real numbers and then it is a (Time Scale), and the question is, what is the form of the angular momentum matrices in different cases of measurement scales?

We can write the mathematical expression which describe how we can see the rotation angles $\theta_i$ in different scales as follow,

$$\mu(\theta_i) = \begin{cases} 0 & l \leq l_{scale} \\ l & l > l_{scale} \end{cases}, i = 1,2,3. \tag{1}$$

## Infinitesimal Generators of the Rotation Symmetry Group $SO(3)$ and Angular Momentum Matrices in different measurement scales

**First case**: $l \leq l_{scale}$

In this case, we have $\mu(\theta_i) = 0, i = 1,2,3$.
$SO(3)$ is the group of all rotations in three dimensions $(x^1, x^2, x^3)$, so if we define $\theta_i$ as the rotation angle around $x^i$ then we can write the representations of the group rotations around $x^1, x^2, x^3$ as

$$R_1(\theta_1) = \begin{pmatrix} 1 & 0 & 0 \\ 0 & cos\theta_1 & -sin\theta_1 \\ 0 & sin\theta_1 & cos\theta_1 \end{pmatrix}, R_2(\theta_2) = \begin{pmatrix} cos\theta_2 & 0 & sin\theta_2 \\ 0 & 1 & 0 \\ -sin\theta_2 & 0 & cos\theta_2 \end{pmatrix},$$

$$R_3(\theta_3) = \begin{pmatrix} cos\theta_3 & -sin\theta_3 & 0 \\ sin\theta_3 & cos\theta_3 & 0 \\ 0 & 0 & 1 \end{pmatrix}. \tag{2}$$

In the first case, we recover our usual picture of continuous rotations and we can find the generators $g_i$ of $R_i(\theta_i)$, which generate the whole Lie group by linearizing each $R_i(\theta_i)$ around $\theta_i = 0$. For example, by using usual Taylor expansion we have for $R_3(\theta_3)$

$$R_3(\theta_3) = I + (\theta_3)\left(\frac{dR_3(\theta_3)}{d\theta_3}\right)_{\theta_3=0} + \left(\frac{\theta_3}{2!}\right)^2 \left(\frac{d^2R_3(\theta_3)}{d\theta_3^2}\right)_{\theta_3=0} + \cdots \qquad (3)$$

where $I$ is the identity matrix.

We have $\left(\frac{dR_3(\theta_3)}{d\theta_3}\right)_{\theta_3=0} = \begin{pmatrix} 0 & -1 & 0 \\ 1 & 0 & 0 \\ 0 & 0 & 0 \end{pmatrix} = g_3$. $\qquad (4)$

$g_3$ is the generator of $R_3(\theta_3)$ and obey the following relations

$$\left(\frac{d^n R_3(\theta_3)}{d\theta_3^n}\right)_{\theta_3=0} = g_3^n. \qquad (5)$$

Similarly, we can write the generators $g_1, g_2$ for the representations $R_1(\theta_1)$, $R_2(\theta_2)$ respectively as

$$g_1 = \begin{pmatrix} 0 & 0 & 0 \\ 0 & 0 & -1 \\ 0 & 1 & 0 \end{pmatrix}, g_2 = \begin{pmatrix} 0 & 0 & 1 \\ 0 & 0 & 0 \\ -1 & 0 & 0 \end{pmatrix}. \qquad (6)$$

Defining $L_i = ig_i$, we arrive at

$$L_1 = \begin{pmatrix} 0 & 0 & 0 \\ 0 & 0 & -i \\ 0 & i & 0 \end{pmatrix}, L_2 = \begin{pmatrix} 0 & 0 & i \\ 0 & 0 & 0 \\ -i & 0 & 0 \end{pmatrix}, L_3 = \begin{pmatrix} 0 & -i & 0 \\ i & 0 & 0 \\ 0 & 0 & 0 \end{pmatrix}. \qquad (7)$$

The matrices $L_1, L_2, L_3$ are the angular momentum matrices (in the context of quantum mechanics) and form a Lie algebra defined by the relation

$[L_i, L_j] = i\epsilon_{ijk}L_k$, $\epsilon_{ijk}$ is the Levi-Civita tensor.

With the above matrices (which is the elements of Lie algebra), we can generate the representations of the entire Lie group $SO(3)$ as

$$R(\theta_1, \theta_2, \theta_3) = e^{-i(\theta_1 L_1 + \theta_2 L_2 + \theta_3 L_3)}, \qquad (8)$$

where $e^{A\theta} = \sum_{m=0}^{\infty} \frac{1}{m!}(A\theta)^m$ for any square matrix $A$.

The previous discussion is the mathematical expression to the following phrase, (The bases of the tangent space of the Lie-group $SO(3)$ at its identity form its associate Lie-algebra $o(3)$ ). Notice that for each Lie-group there is a unique associate Lie-algebra but the inverse is not true.

**Second case:** $l > l_{scale}$

In this case we have $\mu(\theta_i) = l, i = 1,2,3$ and the discretization nature of angles will appear.

Consider again the representation of the rotation around $x^3$

$R_3(\theta_3) = \begin{pmatrix} \cos\theta_3 & -\sin\theta_3 & 0 \\ \sin\theta_3 & \cos\theta_3 & 0 \\ 0 & 0 & 1 \end{pmatrix}$. Generalized Taylor expansion of $R_3(\theta_3)$ on arbitrary structure of space [17] (i.e. on arbitrary time scale) take the form,

$$R_3(\theta_3) = P_n(\theta_3) + E_n(\theta_3), \tag{9}$$

where

$$P_n(\theta_3) = \sum_{k=0}^{n} R_3^{\Delta k}(\theta_3) h_k(\theta_3, s), \tag{10}$$

$$E_n(\theta_3) = \int_s^{\theta_3} h_n(\theta_3, \sigma(\tau)) R^{\Delta(n+1)} \Delta\tau, \tag{11}$$

and $h_k(\theta_3, s)$ defined recursively as follow

$$h_0(\theta_3, s) = 1 \text{ for all } \theta_3, s \in T, \tag{12}$$

and given $h_k$ for $k \in N_0$, we have

$$h_{k+1}(\theta_3, s) = \int_s^{\theta_3} h_k(\tau, s) \Delta\tau \text{ for all } \theta_3, s \in T \tag{13}$$

In our choice of (quantized $\theta_3$), we can take $s = 0$ and then the functions $h_k(\theta_3, 0)$ takes the following forms [ Eq (2) in [17]]

$h_0(\theta_3, 0) = 1$ , $h_k(\theta_3, 0) = \frac{\prod_{r=0}^{k-1}(\theta_3 - rl)}{k!}$ for all $k \geq 1$ and for all $\theta_3 > 0$. (14)

i.e. $h_0(\theta_3, 0) = 1, h_1(\theta_3, 0) = \theta_3, h_2(\theta_3, 0) = \frac{\theta_3(\theta_3 - l)}{2!}$ and so on. (15)

So, the linearized of $R_3(\theta_3)$ around $\theta_3 = 0$ take the form

$$R_3(\theta_3) = I + (\theta_3)\left(\frac{\Delta R_3(\theta_3)}{\Delta \theta_3}\right)_{\theta_3=0} + \frac{\theta_3(\theta_3-l)}{2!}\left(\frac{\Delta^2 R_3(\theta_3)}{\Delta \theta_3^2}\right)_{\theta_3=0} + \cdots , \quad (16)$$

where $\frac{\Delta^n R_3(\theta_3)}{\Delta \theta_3^n}$ is the Hilger (delta) derivative of the order $n$ on the Time Scale $\boldsymbol{T}$.

In our choice, $\theta \in \boldsymbol{T} = \{0, l, 2l, \ldots\ldots, Nl\}$, and the Hilger derivative of any $f(\theta_3)$ with respect to $\theta_3$ take the form

$$\frac{\Delta f(\theta_3)}{\Delta \theta_3} = \frac{f(\sigma(\theta_3))-f(\theta_3)}{\sigma(\theta_3)-\theta_3} = \frac{f(\theta_3+l)-f(\theta_3)}{l}, \quad (17)$$

(Notice that $\sigma(\theta_3) = \theta_3 + l$ for any $\theta_3 \in T - \{2\pi\}$ ).

From (17) we have

$$\frac{\Delta R_3(\theta_3)}{\Delta \theta_3} = \frac{1}{l}\begin{pmatrix} \cos(\theta_3+l)-\cos\theta_3 & -(\sin(\theta_3+l)-\sin\theta_3) & 0 \\ (\sin(\theta_3+l)-\sin\theta_3) & \cos(\theta_3+l)-\cos\theta_3 & 0 \\ 0 & 0 & 0 \end{pmatrix} \quad (18)$$

$$\left(\frac{\Delta R_3(\theta_3)}{\Delta \theta_3}\right)_{\theta_3=0} = \frac{1}{l}\begin{pmatrix} \cos(l)-1 & -\sin(l) & 0 \\ \sin(l) & \cos(l)-1 & 0 \\ 0 & 0 & 0 \end{pmatrix} = A_3. \quad (19)$$

$A_3$ is the generator of the representation of $R_3(\theta_3)$ in the case of quantized rotations, and we can easily verify the following relations

$$\left(\frac{\Delta^n R_3(\theta_3)}{\Delta \theta_3^n}\right)_{\theta_3=0} = A_3^n \quad (20)$$

Similarly, we can find the generators $A_1, A_2$ of the representations $R_1(\theta_1), R_2(\theta_2)$ as follow

$$A_1 = \frac{1}{l}\begin{pmatrix} 0 & 0 & 0 \\ 0 & \cos(l)-1 & -\sin(l) \\ 0 & \sin(l) & \cos(l)-1 \end{pmatrix} \quad (21)$$

$$A_2 = \frac{1}{l}\begin{pmatrix} \cos(l)-1 & 0 & \sin(l) \\ 0 & 0 & 0 \\ -\sin(l) & 0 & \cos(l)-1 \end{pmatrix} \quad (22)$$

Defining $L_i = iA_i$, $\alpha = \frac{\cos(l)-1}{l}$, $\beta = \frac{\sin(l)}{l}$ we finally arrived at the form of the angular momentum matrices in the case of quantized rotations as follow

$$L_1 = \begin{pmatrix} 0 & 0 & 0 \\ 0 & i\alpha & -i\beta \\ 0 & i\beta & i\alpha \end{pmatrix}, L_2 = \begin{pmatrix} i\alpha & 0 & i\beta \\ 0 & 0 & 0 \\ -i\beta & 0 & i\alpha \end{pmatrix}, L_3 = \begin{pmatrix} i\alpha & -i\beta & 0 \\ i\beta & i\alpha & 0 \\ 0 & 0 & 0 \end{pmatrix}. \quad (23)$$

Notice that when $l \to 0$, the angular momentum matrices in (23) take the same form of our usual angular momentum matrices (6).

The relation between Lie group and its associate Lie algebra in arbitrary form of Time Scale was discussed by Hilger in [18] via the following general definition,

Consider a Lie matrix group $GL(n.R)$ and its associate Lie algebra $gl(n.R)$, then the element of a Lie algebra $A$ is an infinitesimal generator of one parameter subgroup $\{y(t): t \in T\} \subseteq GL(n.R)$, where $y(t)$ is the solution of the following matrix initial value problem,

$$y^\Delta = A.y, \quad y(\tau) = I \text{ (identity matrix)}, \quad (24)$$

where $y^\Delta$ is the Hilger derivative and $\tau \in T$ is fixed. For $\tau = 0$, it is easy to verify that the generators $A_1, A_2, A_3$ defined in (19), (21), (22) respectively satisfy the equation (24) for $T = \{0, l, 2l, \ldots\ldots, Nl\}$. i.e. we have the following equations,

$$\frac{\Delta R_i(\theta_i)}{\Delta \theta_i} = A_i. R_i(\theta_i), \quad R_i(\theta_i = 0) = I, \quad i = 1,2,3. \quad (25)$$

Notice that for the angular momentum matrices in the case of quantized rotations,

$$L_1 = \begin{pmatrix} 0 & 0 & 0 \\ 0 & i\alpha & -i\beta \\ 0 & i\beta & i\alpha \end{pmatrix}, L_2 = \begin{pmatrix} i\alpha & 0 & i\beta \\ 0 & 0 & 0 \\ -i\beta & 0 & i\alpha \end{pmatrix} \text{ and } L_3 = \begin{pmatrix} i\alpha & -i\beta & 0 \\ i\beta & i\alpha & 0 \\ 0 & 0 & 0 \end{pmatrix}$$

form a Lie- algebra via generalized definition of the Lie-bracket defined by Hilger in [18].

**Angular momentum operators with observed quantized rotations ($l > l_{scale}$)**

Consider a quantized infinitesimal counter clockwise rotation $\theta$ around $z$ axis, then we have the transformations,

$$\zeta = x\cos\theta - y\sin\theta, \quad \upsilon = x\cos\theta - y\sin\theta, \quad \lambda = z \tag{26}$$

Where $(\zeta, \upsilon, \lambda)$ are the new coordinates.

We can again apply Taylor expansion to get,

$$\zeta = x + \left(\frac{x(\cos l - 1)}{l} - \frac{y\sin l}{l}\right)\theta + O(\theta^2),$$

$$\upsilon = y + \left(\frac{x\sin l}{l} + \frac{y(\cos l - 1)}{l}\right)\theta + O(\theta^2), \quad \lambda = z. \tag{27}$$

As before, let $\alpha = \frac{(\cos l - 1)}{l}, \beta = \frac{\sin l}{l}$, then we can write (27) as follow,

$$\zeta = x + (\alpha x - \beta y)\theta + O(\theta^2),$$

$$\upsilon = y + (\beta x + \alpha y)\theta + O(\theta^2), \quad \lambda = z. \tag{28}$$

Consider some analytical function $(\zeta, \nu, \lambda)$, then we can write

$$f(\zeta, \nu, \lambda) = f(x, y, z) + \left[\alpha\left(x\frac{\partial}{\partial x} + y\frac{\partial}{\partial y}\right) + \beta\left(x\frac{\partial}{\partial y} - y\frac{\partial}{\partial x}\right)\right]f\theta + \cdots \tag{29}$$

Notice that one can making another step and considering the quantization not only of the rotation angles $\theta_i$ but also of the Space itself. i.e. quantization of $(x, y, z)$, and in this case the usual partial derivatives should be replaced by the generalized partial derivatives defined by Bohner and Guseinov in [19].

Now, from (29) consider the Operator,

$$U_3 = \left[\alpha\left(x\frac{\partial}{\partial x} + y\frac{\partial}{\partial y}\right) + \beta\left(x\frac{\partial}{\partial y} - y\frac{\partial}{\partial x}\right)\right]. \tag{30}$$

Similarly, for a quantized infinitesimal counter clockwise rotations around $x$ and $y$ axis we have,

$$U_2 = \left[\alpha\left(z\frac{\partial}{\partial z} + x\frac{\partial}{\partial x}\right) + \beta\left(z\frac{\partial}{\partial x} - y\frac{\partial}{\partial z}\right)\right], \tag{31}$$

$$U_1 = \left[\alpha\left(y\frac{\partial}{\partial y} + z\frac{\partial}{\partial z}\right) + \beta\left(y\frac{\partial}{\partial z} - z\frac{\partial}{\partial y}\right)\right]. \tag{32}$$

Defining $L_i = -ihU_i$, we get the angular momentum operators when the rotation angles are quantized.

Notice that, for example, the angular momentum operator $L_1$ with quantized rotations take the form,

$$L_1 = -ih\left[\alpha\left(y\frac{\partial}{\partial y} + z\frac{\partial}{\partial z}\right) + \beta\left(y\frac{\partial}{\partial z} - z\frac{\partial}{\partial y}\right)\right], \text{ and}$$

$$\lim_{l \to 0} L_1 = -ih\left(y\frac{\partial}{\partial z} - z\frac{\partial}{\partial y}\right).$$ i.e. the same usual form of the Angular momentum Operators. The appearance of the new term $-ih\alpha\left(y\frac{\partial}{\partial y} + z\frac{\partial}{\partial z}\right)$ in $L_1$ and similarly other new terms in $L_2$ and $L_3$, may be an indication of the need to change all quantum mechanical operators in a very small scale (High Energy).

**Quantum magnetic number in low scale**

As we has been showed, if we assume that we can notice the discrete nature of the rotation angles then the angular momentum operators takes the form,

$$L_x = -ih\beta\left(y\frac{\partial}{\partial z} - z\frac{\partial}{\partial y}\right) - ih\alpha\left(y\frac{\partial}{\partial y} + z\frac{\partial}{\partial z}\right),$$

$$L_y = -ih\beta\left(z\frac{\partial}{\partial x} - x\frac{\partial}{\partial z}\right) - ih\alpha\left(z\frac{\partial}{\partial z} + x\frac{\partial}{\partial x}\right),$$

$$L_z = -ih\beta\left(x\frac{\partial}{\partial y} - y\frac{\partial}{\partial x}\right) - ih\alpha\left(x\frac{\partial}{\partial x} + y\frac{\partial}{\partial y}\right). \tag{33}$$

Notice that when $l \to 0$ then $\beta = 1, \alpha = 0$ and $\mu(\theta_i) = 0$ for $i = 1,2,3$. Therefore we recover the continuous structure of the rotation angles and the corresponding usual angular momentum operators.

In the spherical coordinate system with the transformations

$x = r\sin\theta\cos\theta,$

$y = r\sin\theta\sin\varphi,$

$z = r\cos\theta,$  (34)

we can write,

$$L_z = -ih\beta \frac{\partial}{\partial \varphi} - ih\alpha r \sin^2\theta \frac{\partial}{\partial r} - ih\alpha \sin\theta \cos\theta \frac{\partial}{\partial \theta}. \tag{35}$$

**Case (1), measurement in high scale:**

if $l \leq l_{scale}$, then we have $\mu(\theta_i) = 0, \beta = 1, \alpha = 0$, and we recover the usual form of $L_z$ in spherical coordinate system as,

$L_z = -ih \frac{\partial}{\partial \varphi}$. The eigenfunctions of this operator take the form,

$\psi(\varphi) = Ce^{\frac{im\varphi}{h}}$, $C$ is some constant and $m$ is the eigenvalue of the operator $L_z$ which take the following values,

$m = 0, h, 2h, \ldots$ ($m$ is called the magnetic quantum number.)

**Case (2), measurement in low scale**

if $l > l_{scale}$, then we have $\mu(\theta_i) \neq 0$ and the eigenfunctions of the operator $L_z$ in this case can be in general $\psi = \psi(r, \theta, \varphi)$. However, if we can consider a special form of the eigenfunction as a function of $\varphi$ only then we have,

$$-ih\beta \frac{\partial \psi(\varphi)}{\partial \varphi} = m\psi(\varphi). \tag{36}$$

The solution of the last equation take the form,

$\psi(\varphi) = Ce^{\frac{im\varphi}{\beta h}}$, $C$ is some constant.

From the condition $\psi(\varphi) = \psi(\varphi + 2\pi)$, we can write the form of the eigenvalues as,

$m = 0, h\beta, 2h\beta, \ldots$

Therefore, if our scale of measurement $l_{scale}$ is smaller than $l = \frac{2\pi}{N}$, then the quantum magnetic number shifted to fractional multiples of $h$. A similar argument show that the spin also take some fractional values between integer and half integer and therefore the quantum particles in this low scale maybe obey some intermediate statistics between Fermions and Bosons. For a possible form of intermediate statistics between Fermions and bosons see [20].

## Conclusion

The angular momentum matrices and operators with quantized rotations have been derived by using Time-Scale Calculus. The values of quantum magnetic number in low scale have been calculated. Finally, if we want to study physical theories on more general structures of usual continuous spacetime, then the price is using non usual mathematical frame work called time-scale calculus.